\documentclass[preprint,showpacs,showkeys,superscriptaddress,nofootinbib]{revtex4}
  \usepackage{graphicx}
  \usepackage{rotating}
  \usepackage{multirow}
  \begin{document}
  \title{Study on the ${\Upsilon}(1S)$ ${\to}$ $B_{c}M$ weak decays}
  \author{Junfeng Sun}
  \affiliation{Institute of Particle and Nuclear Physics,
              Henan Normal University, Xinxiang 453007, China}
  \author{Lili Chen}
  \affiliation{Institute of Particle and Nuclear Physics,
              Henan Normal University, Xinxiang 453007, China}
  \author{Na Wang}
  \affiliation{Institute of Particle and Key Laboratory of Quark and Lepton Physics,
              Central China Normal University, Wuhan 430079, China}
  \affiliation{Institute of Particle and Nuclear Physics,
              Henan Normal University, Xinxiang 453007, China}
  \author{Qin Chang}
  \thanks{corresponding author}
  \affiliation{Institute of Particle and Nuclear Physics,
              Henan Normal University, Xinxiang 453007, China}
  \affiliation{Institute of Particle and Key Laboratory of Quark and Lepton Physics,
              Central China Normal University, Wuhan 430079, China}
  \author{Jinshu Huang}
  \affiliation{College of Physics and Electronic Engineering,
              Nanyang Normal University, Nanyang 473061, China}
  \affiliation{Institute of Particle and Nuclear Physics,
              Henan Normal University, Xinxiang 453007, China}
  \author{Yueling Yang}
  \affiliation{Institute of Particle and Nuclear Physics,
              Henan Normal University, Xinxiang 453007, China}

  \begin{abstract}
  Motivated by the prospects of the potential ${\Upsilon}(1S)$ particle
  at high-luminosity heavy-flavor experiments, we studied the
  ${\Upsilon}(1S)$ ${\to}$ $B_{c}M$ weak decays, where $M$ $=$ ${\pi}$,
  ${\rho}$, $K^{(\ast)}$.
  The nonfactorizable contributions to hadronic matrix elements
  are taken into consideration with the QCDF approach.
  It is found that the CKM-favored ${\Upsilon}(1S)$ ${\to}$
  $B_{c}{\rho}$ decay has branching ratio of ${\cal O}(10^{-10})$,
  which might be measured promisingly by the future experiments.
  \end{abstract}
  \maketitle

  \section{Introduction}
  \label{sec01}
  The first evidence for upsilons, the bound states of $b\bar{b}$,
  was observed in collisions of protons with a stationary nuclear
  target at Fermilab in 1977 \cite{herb,innes}.
  From that moment on, bottomonia have been a subject of intensive
  experimental and theoretical research.
  Some of the salient features of upsilons are as follows \cite{ann1983}:
  (1)
  In the upsilons rest frame, the relative motion
  of the $b$ quark is sufficiently slow.
  Nonrelativistic Schr\"{o}dinger equation can be used to
  describe the spectrum of bottomonia states and
  thus one can learn about the interquark binding
  forces.
  (2)
  The ${\Upsilon}(nS)$ particles, with the radial
  quantum number $n$ $=$ $1$, $2$ and $3$,
  decay primarily via the annihilation of the $b\bar{b}$
  quark pairs into three gluons.
  Thus the properties of the invisible gluons and of the
  gluon-quark coupling can be gleaned through the study
  of the upsilons decay.
  (3)
  Compared with the $u$, $d$, $s$ light quarks,
  the relatively large mass of the heavy $b$ quark
  implies a nonnegligible coupling to the Higgs bosons,
  making upsilons to be one of the best hunting
  grounds for light Higgs particles.
  By now, our knowledge of the properties of bottomonia comes
  primarily from $e^{+}e^{-}$ annihilation.

  The ${\Upsilon}(1S)$ particle is the ground state
  of the vector bottomonia with quantum number of
  $I^{G}J^{PC}$ $=$ $0^{-}1^{--}$ \cite{pdg}.
  The mass of the ${\Upsilon}(1S)$ particle,
  $m_{{\Upsilon}(1S)}$ $=$ $9460.30{\pm}0.26$ \cite{pdg},
  is about three times heavier than the mass of the
  $J/{\psi}$ particle (the ground state of charmonia with
  the same quantum number of $I^{G}J^{PC}$).
  On one hand,
  compared with the $J/{\psi}$ decay, much richer decay channels
  could be accessed by the ${\Upsilon}(1S)$ particle.
  On the other hand,
  the coupling constant ${\alpha}_{s}$ for the ${\Upsilon}(1S)$
  decay is smaller than that for the $J/{\psi}$ decay due to
  the QCD nature of asymptotic freedom, which results in
  hadronic partial width ${\Gamma}({\Upsilon}(1S){\to}ggg)$
  $<$ ${\Gamma}(J/{\psi}{\to}ggg)$, although the possible
  phase space in the ${\Upsilon}(1S)$ decay
  is larger than that in the $J/{\psi}$ decay.
  In addition, the squared value of the $b$ quark charge,
  $Q^{2}_{b}$ $=$ $1/9$,
  is less than that of the $c$ quark charge,
  $Q^{2}_{c}$ $=$ $4/9$,
  which results in electromagnetic partial width
  ${\Gamma}({\Upsilon}(1S){\to}{\gamma}^{\ast})$
  $<$ ${\Gamma}(J/{\psi}{\to}{\gamma}^{\ast})$.
  So the full decay width of the ${\Upsilon}(1S)$ particle,
  ${\Gamma}_{{\Upsilon}(1S)}$ $=$ $54.02{\pm}1.25$ keV,
  is less than that of the $J/{\psi}$ particle,
  ${\Gamma}_{J/{\psi}}$ $=$ $92.9{\pm}2.8$ keV \cite{pdg}.
  Furthermore, one of the outstanding properties of all
  upsilons below $B\bar{B}$ threshold is their narrow decay
  width of tens of keV \cite{pdg}.

  The ${\Upsilon}(1S)$ and $J/{\psi}$ particles share the
  similar decay mechanism.
  As is the case for the $J/{\psi}$ particle, strong decays of the
  ${\Upsilon}(1S)$ particle are suppressed by the phenomenological
  OZI (Okubo-Zweig-Iizuka) rules \cite{ozi-o,ozi-z,ozi-i},
  so electromagnetic interactions and radiative transitions
  become competitive.
  It is expected that, at the lowest order approximation,
  the decay modes of the ${\Upsilon}(1S)$ particle could be
  subdivided into four types:
  (1) The lion's share of the decay width is the hadronic decay
  via the annihilation of the $b\bar{b}$ quark pairs into three
  gluons, i.e., some $(81.7{\pm}0.7)\%$ via ${\Upsilon}(1S)$
  ${\to}$ $ggg$ \cite{pdg}.
  (2) The partial width of the electromagnetic decay via the
  annihilation of the $b\bar{b}$ quark pairs into a virtual
  photon could be written approximately as
  $(3+R){\Gamma}_{{\ell}{\ell}}$,
  where the value of $R$ is the ratio of inclusive production
  of hadrons to the ${\mu}^{+}{\mu}^{-}$ pair production rate
  at the energy scale of $m_{{\Upsilon}(1S)}$,
  and ${\Gamma}_{{\ell}{\ell}}$ is the partial width of decay
  into dileptons.
  (3) Branching ratio of the radiative decay is about
  $Br({\Upsilon}(1S){\to}{\gamma}gg)$
  ${\simeq}$ $(2.2{\pm}0.6)\%$ \cite{pdg}.
  The up-to-date research for light Higgs bosons in the
  ${\Upsilon}(1S)$ radiative decay has been performed
  by CLEO \cite{0807.1427}, Belle\cite{pos2013.higgs}
  and BaBar \cite{1502.06019} Collaborations.
  (4) The magnetic dipole transition decay,
  ${\Upsilon}(1S)$ ${\to}$ ${\gamma}{\eta}_{b}(1S)$,
  is very challenging to experimental physicist
  due to the very soft photon and pollution from
  other processes, such as ${\Upsilon}(1S)$ ${\to}$
  ${\pi}^{0}X$ ${\to}$ ${\gamma}{\gamma}X$
  \cite{1212.6552}.
  The experimental signal for ${\Upsilon}(1S)$ ${\to}$
  ${\gamma}{\eta}_{b}(1S)$ has not been discovered
  until now.
  Besides, the ${\Upsilon}(1S)$ particle could also decay
  via the weak interactions, although the branching
  ratio for a single $b$ or $\bar{b}$ quark decay is
  tiny, about $2/{\tau}_{B}{\Gamma}_{{\Upsilon}(1S)}$
  ${\sim}$ $10^{-8}$ \cite{pdg}.
  In this paper, we will estimate the branching ratios
  for the flavor-changing nonleptonic ${\Upsilon}(1S)$
  ${\to}$ $B_{c}M$ weak decays with the QCD factorization
  (QCDF) approach \cite{qcdf1,qcdf2},
  where $M$ $=$ ${\pi}$, ${\rho}$, $K$ and $K^{\ast}$.
  The motivation is listed as follows.

   \begin{table}[h]
   \caption{Summary of data samples (in the unit of $10^{6}$)
   of the ${\Upsilon}(nS)$ particles below $B\bar{B}$ threshold
   collected by Belle, BaBar and CLEO Collaborations.
   The data in the 5th column correspond to total number
   of the ${\Upsilon}(1S)$ particle, including events from
   hadronic dipion transitions between upsilons
   ${\Upsilon}(2S,3S)$ ${\to}$ ${\pi}{\pi}{\Upsilon}(1S)$,
   where branching ratios for the dipion decays
   ${\Upsilon}(2S)$ ${\to}$
   ${\pi}^{+}{\pi}^{-}{\Upsilon}(1S)$,
   ${\pi}^{0}{\pi}^{0}{\Upsilon}(1S)$ and
   ${\Upsilon}(3S)$ ${\to}$
   ${\pi}^{+}{\pi}^{-}{\Upsilon}(1S)$,
   ${\pi}^{0}{\pi}^{0}{\Upsilon}(1S)$
   are $(17.85{\pm}0.26)\%$, $(8.6{\pm}0.4)\%$,
   $(4.37{\pm}0.08)\%$ and $(2.20{\pm}0.13)\%$,
   respectively \cite{pdg}.}
   \label{tot}
   \begin{ruledtabular}
   \begin{tabular}{c|c|c|c|c}
    & ${\Upsilon}(1S)$ & ${\Upsilon}(2S)$
    & ${\Upsilon}(3S)$ & total ${\Upsilon}(1S)$ \\ \hline
   Belle \cite{1212.5342} & 102    & 158    & 11    & 145 \\
   BaBar \cite{1108.5874} &        & 121.8  & 98.6  &  39 \\
   CLEO  \cite{0704.2766} &  22.78 &   9.45 &  8.89 &  26 \\
   \end{tabular}
   \end{ruledtabular}
   \end{table}

  From the experimental point of view,
  (1)
  there is plenty of upsilons at the high-luminosity
  dedicated bottomonia factories, for example,
  over $10^{8}$ ${\Upsilon}(1S)$ at Belle (see Table \ref{tot}).
  Upsilons are also observed by the on-duty
  ALICE \cite{1403.3648}, ATLAS \cite{1212.7255},
  CMS \cite{1501.07750}, LHCb \cite{1304.6977}
  experiments at LHC.
  It is hopefully expected that more upsilons
  could be accumulated with great precision
  at the running upgraded LHC and forthcoming
  SuperKEKB.
  The huge ${\Upsilon}(1S)$ data samples will provide
  good opportunities to search for the ${\Upsilon}(1S)$
  weak decays which in some cases might be detectable.
  Theoretical studies on the ${\Upsilon}(1S)$
  weak decays are just necessary to offer a ready
  reference.
  (2)
  For the two-body ${\Upsilon}(1S)$ ${\to}$ $B_{c}M$ decay,
  final states with opposite charges have definite
  energies and momenta in the center-of-mass frame
  of the ${\Upsilon}(1S)$ particle.
  Particularly, identification of a single charged
  $B_{c}$ meson in the final state, which is free from
  inefficiently double tagging of the bottomed hadron
  pairs occurring above the $B\bar{B}$ threshold,
  would provide an unambiguous signature of the
  ${\Upsilon}(1S)$ weak decay.
  Of course, small branching ratios make the observation
  of the ${\Upsilon}(1S)$ weak decays extremely difficult,
  and evidences of an abnormally large production rate
  of single $B_{c}$ mesons in the ${\Upsilon}(1S)$ decay
  might be a hint of new physics beyond the standard model.

  From the theoretical point of view,
  the bottom-changing upsilon weak decay could permit
  overconstraining parameters obtained from $B$ meson decay,
  but few studies devoted to the nonleptonic upsilon
  weak decay in the past.
  For the ${\Upsilon}(1S)$ ${\to}$ $B_{c}M$ decay,
  the amplitude is usually treated as the factorizable
  product of two independent factors: one describing
  the transition between heavy quarkonium ${\Upsilon}(1S)$
  and $B_{c}$, and the other depicting the production
  of the $M$ state from the vacuum.
  Previous works, such as Ref. \cite{zpc62.271} based on the
  spin symmetry and nonrecoil approximation,
  Ref. \cite{ijma14} based on the heavy quark effective theory,
  and Ref. \cite{adv2013} based on the Bauer-Stech-Wirbel
  (BSW) model \cite{bsw1}, concentrated mainly upon the
  ${\Upsilon}(1S)$ ${\to}$ $B_{c}$ transition form factors
  which are related
  to the space integrals of the meson wave functions.
  As is well known, there exist hierarchical scales with
  nonrelativistic quantum chromodynamics (NRQCD)
  \cite{prd46,prd51,rmp77} which is an
  approach to deal with the heavy quarkonium, i.e.
  $M^{2}$ ${\gg}$ $(Mv)^{2}$ ${\gg}$ $(Mv^{2})^{2}$,
  where $M$ is the mass of heavy quark with typical
  velocities $v$ ${\sim}$ ${\alpha}_{s}$ ${\ll}$ $1$.
  The physical contributions at scales of $Mv^{2}$ ${\sim}$
  ${\Lambda}_{\rm QCD}$
  are absorbed into wave functions of the ${\Upsilon}(1S)$
  and $B_{c}$ particles, thus the ${\Upsilon}(1S)$
  ${\to}$ $B_{c}$ transition should be dominated by
  the nonperturbative dynamics.
  The nonfactorizable contributions above scales of $Mv^{2}$
  has not be taken seriously in previous works.
  About 2000, M. Beneke {\em et al.} proposed the QCDF
  approach \cite{qcdf1,qcdf2}, where nonfactorizable
  contributions could be estimated systematically with
  the perturbation theory
  based on collinear factorization approximation and
  power countering rules in the heavy quark limit
  \cite{qcdf2},
  and the QCDF approach has been widely applied
  to nonleptonic $B$ meson decays.
  So it should be very interesting to study the ${\Upsilon}(1S)$
  ${\to}$ $B_{c}M$ weak decays by considering nonfactorizable
  contributions with the attractive QCDF approach.

  This paper is organized as follows.
  In section \ref{sec02}, we will present the theoretical framework
  and the amplitudes for the nonleptonic two-body ${\Upsilon}(1S)$
  ${\to}$ $B_{c}M$ weak decays with the QCDF approach.
  Section \ref{sec03} is devoted to numerical results and discussion.
  The last section is our summary.

  \section{theoretical framework}
  \label{sec02}
  \subsection{The effective Hamiltonian}
  \label{sec0201}
  The low energy effective Hamiltonian responsible for the
  ${\Upsilon}(1S)$ ${\to}$ $B_{c}M$ decays is \cite{9512380}
   \begin{equation}
  {\cal H}_{\rm eff}\ =\ \frac{G_{F}}{\sqrt{2}}\,
   \sum\limits_{q=d,s}\, V_{cb} V_{uq}^{\ast}\,
   \Big\{ C_{1}({\mu})\,Q_{1}({\mu})
         +C_{2}({\mu})\,Q_{2}({\mu}) \Big\}
   + {\rm h.c.}
   \label{hamilton},
   \end{equation}
  where the Fermi coupling constant $G_{F}$ $=$
  $1.166{\times}10^{-5}\,{\rm GeV}^{-2}$ \cite{pdg};
  Using the Wolfenstein parameterization, the
  Cabibbo-Kobayashi-Maskawa (CKM) factors can be
  expanded as a power series in the small parameter
  ${\lambda}$ $=$ $0.22537(61)$ \cite{pdg},
  \begin{eqnarray}
  V_{cb}V_{ud}^{\ast} &=&
               A{\lambda}^{2}
  - \frac{1}{2}A{\lambda}^{4}
  - \frac{1}{8}A{\lambda}^{6}
  +{\cal O}({\lambda}^{8})
  \label{eq:ckm01}, \\
  V_{cb}V_{us}^{\ast} &=& A{\lambda}^{3}
  +{\cal O}({\lambda}^{8})
  \label{eq:ckm02}.
  \end{eqnarray}
  The Wilson coefficients $C_{1,2}(\mu)$ summarize the
  physical contributions above scales of ${\mu}$.
  The Wilson coefficients $C_{i}$ are calculable with the
  perturbation theory and have properly been evaluated to
  the next-to-leading order (NLO) with the renormalization
  group (RG) equation.
  The numerical values of the Wilson coefficients $C_{1,2}$
  at scales of ${\mu}$ ${\sim}$
  ${\cal O}(m_{b})$ in naive dimensional regularization
  scheme are listed in Table \ref{tab:ci}.
  The local tree four-quark operators are defined as follows.
    \begin{eqnarray}
    Q_{1} &=&
  [ \bar{c}_{\alpha}{\gamma}_{\mu}(1-{\gamma}_{5})b_{\alpha} ]
  [ \bar{q}_{\beta} {\gamma}^{\mu}(1-{\gamma}_{5})u_{\beta} ]
    \label{q1}, \\
    Q_{2} &=&
  [ \bar{c}_{\alpha}{\gamma}_{\mu}(1-{\gamma}_{5})b_{\beta} ]
  [ \bar{q}_{\beta}{\gamma}^{\mu}(1-{\gamma}_{5})u_{\alpha} ]
    \label{q2},
    \end{eqnarray}
  where ${\alpha}$ and ${\beta}$ are color indices and the
  sum over repeated indices is understood.
  To obtain the decay amplitudes, the remaining and the
  most intricate works are to calculate accurately
  hadronic matrix elements of local operators.

  \subsection{Hadronic matrix elements}
  \label{sec0202}
  Phenomenologically, the simplest treatment with hadronic matrix
  elements of four-quark operators is approximated by the product
  of two current matrix elements with color transparency
  ansatz \cite{bjorken} and naive factorization (NF) scheme
  \cite{nf1,nf2}, and current matrix elements are further
  parameterized by decay constants and transition form factors.
  For example, previous studies on the ${\Upsilon}(1S)$ ${\to}$
  $B_{c}M$ decay \cite{ijma14,adv2013} were based on NF approach.

  As is well known, NF's defect is the disappearance of the
  renormalization scale dependence, the strong phases and the
  nonfactorizatable contributions from hadronic matrix elements,
  resulting in nonphysical amplitudes and no $CP$ violating
  asymmetries.
  To remedy NF's deficiencies, M. Beneke {\em et al.} proposed that
  hadronic matrix elements could be written as the convolution
  integrals of hard scattering kernels and light cone distribution
  amplitudes with the QCDF approach \cite{qcdf1,qcdf2}.

  For the ${\Upsilon}(1S)$ ${\to}$ $B_{c}M$ decay,
  the spectator quark is the heavy bottom (anti)quark.
  According to the QCDF's power counting rules \cite{qcdf2},
  contributions from the spectator scattering are
  power suppressed.
  With the QCDF master formula, hadronic matrix elements
  could be written as :
   \begin{equation}
  {\langle}B_{c}M{\vert}Q_{i}{\vert}{\Upsilon}(1S){\rangle} =
   \sum\limits_{i} F_{i}^{ {\Upsilon}{\to}B_{c} }
  {\int}\,dx\, H_{i}(x)\,{\phi}_{M}(x)
   \label{hadronic},
   \end{equation}
  where transition form factor $F_{i}^{ {\Upsilon}{\to}B_{c} }$
  and light cone distribution amplitudes ${\phi}_{M}(x)$ of the
  emitted meson $M$ are nonperturbative input parameters,
  hard scattering kernels $H_{i}(x)$ are computable order by
  order with the perturbation theory in principle.

  The leading twist two-valence-particle distribution
  amplitudes of pseudoscalar and longitudinally polarized
  vector meson are defined in terms of Gegenbauer
  polynomials \cite{ballv,ballp}:
   \begin{equation}
  {\phi}_{M}(x)=6\,x\bar{x}
   \sum\limits_{n=0}^{\infty}
   a_{n}^{M}\, C_{n}^{3/2}(x-\bar{x})
   \label{twist},
   \end{equation}
  where $\bar{x}$ $=$ $1$ $-$ $x$;
  $a_{n}^{M}$ is the Gegenbauer moment
  and $a_{0}^{M}$ ${\equiv}$ $1$.

  After calculation, the decay amplitudes could be written as
   \begin{equation}
  {\cal A}({\Upsilon}(1S){\to}B_{c}M) =
  {\langle}B_{c}M{\vert}{\cal H}_{\rm eff}{\vert}{\Upsilon}(1S){\rangle} =
   \frac{G_{F}}{\sqrt{2}}\, V_{cb} V_{uq}^{\ast}\, a_{1}\,
  {\langle}M{\vert}J^{\mu}{\vert}0{\rangle}
  {\langle}B_{c}{\vert}J_{\mu}{\vert}{\Upsilon}(1S){\rangle}
   \label{amp}.
   \end{equation}

  The coefficient $a_{1}$ in Eq.(\ref{amp}),
  including nonfactorizable contributions from QCD
  radiative vertex corrections, is written as \cite{prd77}:
  \begin{equation}
   a_{1}
    = C_{1}^{\rm NLO}+\frac{1}{N_{c}}\,C_{2}^{\rm NLO}
    + \frac{{\alpha}_{s}}{4{\pi}}\, \frac{C_{F}}{N_{c}}\,
      C_{2}^{\rm LO}\, V
   \label{a1}.
  \end{equation}

  For the transversely polarized vector meson, the factor $V$
  is zero beyond leading twist (twist-2) contributions.
  For the pseudoscalar and longitudinally polarized vector meson,
  with the modified minimal subtraction ($\overline{\rm MS}$)
  scheme, the factor $V$ is written as \cite{prd77}:
  \begin{equation}
  V = 3\,{\log} \Big( \frac{ m_{b}^{2} }{ {\mu}^{2} } \Big)
    + 3\,{\log} \Big( \frac{ m_{c}^{2} }{ {\mu}^{2} } \Big)
    - 18 +{\int}_{0}^{1}dx\, T(x)\,{\phi}_{M}(x)
  \label{vc01},
  \end{equation}
  where
  \begin{eqnarray}
  T(x) &=&
      \frac{c_{a}}{1-c_{a}}\,{\log}(c_{a})
    - \frac{4\,c_{b}}{1-c_{b}}\,{\log}(c_{b})
      \nonumber \\ &+&
      \frac{c_{d}}{1-c_{d}}\,{\log}(c_{d})
    - \frac{4\,c_{c}}{1-c_{c}}\,{\log}(c_{c})
      \nonumber \\ &-&
      r_{c}\, \Big\{ \frac{c_{a}}{(1-c_{a})^{2}}\,{\log}(c_{a})
    + \frac{1}{1-c_{a}} \Big\}
      \nonumber \\ &-&
      r_{c}^{-1}\,\Big\{ \frac{c_{d}}{(1-c_{d})^{2}}\,{\log}(c_{d})
    + \frac{1}{1-c_{d}} \Big\}
      \nonumber \\ &+&
      f(c_{a})-f(c_{b})-f(c_{c})+f(c_{d})
      \nonumber \\ &+&
      2\, {\log}(r_{c}^{2}) \{ {\log}(c_{a}) -{\log}(c_{b}) \}
  \label{vc02},
  \end{eqnarray}
  \begin{equation}
   f(c)\ =\ 2\,{\rm Li}_{2} \Big( \frac{c-1}{c} \Big)
   -{\log}^{2}(c)-\frac{2\,c}{1-c}{\log}(c),
   \label{ff}
   \end{equation}
  and the relations
  \begin{eqnarray}
  r_{c} &=& m_{c}/m_{b}
  \label{cr}, \\
  c_{a} &=& x\,(1-r_{c}^{2})
  \label{ca}, \\
  c_{b} &=& \bar{x}\,(1-r_{c}^{2})
  \label{cb}, \\
  c_{c} &=& -c_{a}/r_{c}^{2}
  \label{cc}, \\
  c_{d} &=& -c_{b}/r_{c}^{2}
  \label{cd}.
  \end{eqnarray}

  The numerical values of coefficient $a_{1}$
  at scales of ${\mu}$ ${\sim}$
  ${\cal O}(m_{b})$ are listed in Table \ref{tab:ci}.
   \begin{table}[h]
   \caption{The numerical values of the Wilson coefficients $C_{1,2}$
    and $a_{1}$ for the ${\Upsilon}(1S)$ ${\to}$ $B_{c}{\pi}$ decay
    at different scales, where $m_{b}$ $=$ 4.78 GeV \cite{pdg}.}
   \label{tab:ci}
   \begin{ruledtabular}
   \begin{tabular}{c|cc|cc|cc}
 & \multicolumn{2}{c|}{NLO} & \multicolumn{2}{c|}{LO}
 & \multicolumn{2}{c}{QCDF} \\ \cline{2-7}
 ${\mu}$ & $C_{1}$ & $C_{2}$ & $C_{1}$ & $C_{2}$
         & Re($a_{1}$) & Im($a_{1}$) \\ \hline
 $0.5\,m_{b}$ & $ 1.128$ & $-0.269$
              & $ 1.168$ & $-0.337$
              & $ 1.076$ & $+0.027$ \\
    $m_{b}$   & $ 1.076$ & $-0.173$
              & $ 1.110$ & $-0.235$
              & $ 1.054$ & $+0.015$ \\
 $1.5\,m_{b}$ & $ 1.054$ & $-0.128$
              & $ 1.085$ & $-0.188$
              & $ 1.043$ & $+0.011$ \\
 $2.0\,m_{b}$ & $ 1.041$ & $-0.100$
              & $ 1.070$ & $-0.159$
              & $ 1.036$ & $+0.008$
   \end{tabular}
   \end{ruledtabular}
   \end{table}

  \subsection{Decay constants and form factors}
  \label{sec0203}
  The matrix elements of current operators are defined as follows:
   \begin{eqnarray}
  {\langle}P(p){\vert}A_{\mu}{\vert}0{\rangle}
  &=&
   -if_{P}\,p_{\mu}
   \label{cme01}, \\
  {\langle}V(p,{\epsilon}){\vert}V_{\mu}{\vert}0{\rangle}
  &=&
   f_{V}\,m_{V}\,{\epsilon}_{V,{\mu}}^{\ast}
   \label{cme02},
   \end{eqnarray}
 where $f_{P}$ and $f_{V}$ are the decay constants
 of pseudoscalar and vector mesons, respectively;
 $m_{V}$ and ${\epsilon}_{V}$ denote the mass and
 polarization of vector meson, respectively.

  The transition form factors are defined as follows
  \cite{ijma14,adv2013,bsw1}:
    \begin{eqnarray}
   & &
   {\langle}B_{c}(p_{2}){\vert}V_{\mu}-A_{\mu}
   {\vert}{\Upsilon}(p_{1},{\epsilon}){\rangle}
    \nonumber \\ &=&
  -{\epsilon}_{{\mu}{\nu}{\alpha}{\beta}}\,
   {\epsilon}_{{\Upsilon}}^{{\nu}}\,
    q^{\alpha}\, (p_{1}+p_{2})^{\beta}\,
     \frac{V^{{\Upsilon}{\to}B_{c}}(q^{2})}{m_{{\Upsilon}}+m_{B_{c}}}
   -i\,\frac{2\,m_{{\Upsilon}}\,{\epsilon}_{{\Upsilon}}{\cdot}q}{q^{2}}\,
    q_{\mu}\, A_{0}^{{\Upsilon}{\to}B_{c}}(q^{2})
    \nonumber \\ & &
    -i\,{\epsilon}_{{\Upsilon},{\mu}}\,
    ( m_{{\Upsilon}}+m_{B_{c}} )\, A_{1}^{{\Upsilon}{\to}B_{c}}(q^{2})
   -i\,\frac{{\epsilon}_{{\Upsilon}}{\cdot}q}{m_{{\Upsilon}}+m_{B_{c}}}\,
   ( p_{1} + p_{2} )_{\mu}\, A_{2}^{{\Upsilon}{\to}B_{c}}(q^{2})
    \nonumber \\ & &
   +i\,\frac{2\,m_{{\Upsilon}}\,{\epsilon}_{{\Upsilon}}{\cdot}q}{q^{2}}\,
   q_{\mu}\, A_{3}^{{\Upsilon}{\to}B_{c}}(q^{2})
    \label{cme03},
    \end{eqnarray}
  where $q$ $=$ $p_{1}$ $-$ $p_{2}$;
  and $A_{0}(0)$ $=$ $A_{3}(0)$
  is required compulsorily to cancel singularities at the
  pole $q^{2}$ $=$ $0$.
  There is a relation among these form factors
  \begin{equation}
   2m_{{\Upsilon}}A_{3}(q^{2})=
   (m_{{\Upsilon}}+m_{B_{c}})A_{1}(q^{2})
  +(m_{{\Upsilon}}-m_{B_{c}})A_{2}(q^{2})
  \label{form01}.
  \end{equation}

  It is clearly seen that there are only three independent
  form factors, $A_{0,1}(0)$ and $V(0)$, at the pole
  $q^{2}$ $=$ $0$ for the ${\Upsilon}(1S)$ ${\to}$ $B_{c}M$
  decays.
  The form factors at the pole $q^{2}$ $=$ $0$ could be
  written as the overlap integrals of wave functions of mesons
  \cite{bsw1}, i.e.,
  \begin{equation}
  A_{0}^{{\Upsilon}{\to}B_{c}}(0) =
  {\int}d\vec{k}_{\perp} {\int}_{0}^{1}dx\,
   \Big\{ {\Phi}_{\Upsilon}(k_{\perp},x,1,0)\,
  {\sigma}_{z}\, {\Phi}_{B_{c}}(k_{\perp},x,0,0) \Big\}
  \label{form-a0},
  \end{equation}
  \begin{eqnarray}
  A_{1}^{ {\Upsilon}{\to}B_{c} }(0) &=&
  \frac{ m_{b}+m_{c} }{ m_{{\Upsilon}(1S)}+m_{B_{c}} }
  I^{{\Upsilon}{\to}B_{c}}
  \label{form-a1} \\
  V^{ {\Upsilon}{\to}B_{c} }(0)  &=&
  \frac{ m_{b}-m_{c} }{ m_{{\Upsilon}(1S)}-m_{B_{c}} }
  I^{{\Upsilon}{\to}B_{c}}
  \label{form-v},
  \end{eqnarray}
  \begin{equation}
  I^{{\Upsilon}{\to}B_{c}} = \sqrt{2}
  {\int}d\vec{k}_{\perp} {\int}_{0}^{1} \frac{dx}{x}\,
   \Big\{ {\Phi}_{\Upsilon}(k_{\perp},x,1,-1)\,
  i{\sigma}_{y}\, {\Phi}_{B_{c}}(k_{\perp},x,0,0) \Big\}
  \label{form-ii},
  \end{equation}
  where ${\sigma}_{y,z}$ is a Pauli matrix acting on
  the spin indices of the decaying bottom quark;
  $x$ and $\vec{k}_{\perp}$ denote the fraction of
  the longitudinal momentum and the transverse momentum
  carried by the nonspectator quark, respectively.

  Using the separation of momentum and spin variables,
  the wave functions of mesons can be written as
  \begin{equation}
  {\Phi}(\vec{k}_{\perp},x,j,j_{z})\, =
  {\phi}(\vec{k}_{\perp},x)\,{\vert}s,s_{z},s_{1},s_{2}{\rangle}
  \label{wave01},
  \end{equation}
  with the normalization condition,
  \begin{equation}
  \sum\limits_{s_{1},s_{2}} {\int}d\vec{k}_{\perp}
  {\int}_{0}^{1}dx\, {\vert}{\Phi}(\vec{k}_{\perp},x,j,j_{z})
  {\vert}^{2}\, = 1
  \label{wave02},
  \end{equation}
  where $s_{1,2}$ denote the spin of valence quark in meson;
  $\vec{s}$ $=$ $\vec{s}_{1}$ $+$ $\vec{s}_{2}$;
  $s$ $=$ $1$ and $0$ for the ${\Upsilon}(1S)$ and
  $B_{c}$ particles, respectively.

  For the ground states of heavy quarkonia ${\Upsilon}(1S)$ and $B_{c}$
  particles, we will take the solution of the Sch\"{o}dinger
  equation with nonrelativistic three-dimensional scalar harmonic
  oscillator potential,
   \begin{equation}
  {\phi}(\vec{k})\ {\sim}\
  {\int}d\vec{r}\,{\phi}(\vec{r})\,e^{-i\vec{k}{\cdot}\vec{r}}\ {\sim}\
  {\int}d\vec{r}\,e^{-{\beta}^{2}r^{2}/2} \,e^{-i\vec{k}{\cdot}\vec{r}}\
  {\sim}\ e^{-\vec{k}^{2}/2{\beta}^{2}}
   \label{wave03},
   \end{equation}
  where the parameter ${\beta}$ determines the average
  transverse quark momentum, i.e.,
  ${\langle}\vec{k}^{2}_{\perp}{\rangle}$ $=$ ${\beta}^{2}$.
  According to the power counting rules of NRQCD \cite{prd46},
  the characteristic magnitude of the moment is order of
  $Mv$ and $v$ ${\sim}$ ${\alpha}_{s}$.
  So we will take ${\beta}$ $=$ $M{\alpha}_{s}$ in our
  calculation.
  Employing the substitution ansatz \cite{xiao},
   \begin{equation}
   \vec{k}^{2}\ {\to}\ \frac{1}{4} \Big\{
   \frac{\vec{k}_{\perp}^{2}+m_{1}^{2}}{x_{1}}
  +\frac{\vec{k}_{\perp}^{2}+m_{2}^{2}}{x_{2}}
  +\frac{(m_{1}^{2}-m_{2}^{2})^{2}}{
   \frac{\vec{k}_{\perp}^{2}+m_{1}^{2}}{x_{1}}
  +\frac{\vec{k}_{\perp}^{2}+m_{2}^{2}}{x_{2}} }
   \Big\}
   \label{wave04},
   \end{equation}
   where $x_{1}$ $+$ $x_{2}$ $=$ $1$;
   $m_{1,2}$ is the mass of valence quark.
   Setting $x_{1}$ $=$ $x$, we can obtain
   \begin{equation}
   {\phi}(\vec{k}_{\perp},x)\ = N
   {\exp}\Big\{ \frac{-1}{ 8{\beta}^{2} } \Big[
   \frac{\vec{k}_{\perp}^{2}+\bar{x}\,m_{1}^{2}+x\,m_{2}^{2} }{ x\,\bar{x} }
  +\frac{ (m_{1}^{2}-m_{2}^{2})^{2}\,x\,\bar{x} }{
   \vec{k}_{\perp}^{2}+\bar{x}\,m_{1}^{2}+x\,m_{2}^{2} }
   \Big] \Big\}
   \label{wave04},
   \end{equation}
  where $N$ is a normalization factor.

  Using the aforementioned convention, we get
   \begin{eqnarray}
   &
   A_{0}^{{\Upsilon}{\to}B_{c}}(0)\, =\,
   A_{3}^{{\Upsilon}{\to}B_{c}}(0)\, =\,
   0.81{\pm}0.01,
   &  \label{wave05} \\
   &
   A_{1}^{{\Upsilon}{\to}B_{c}}(0)\, =\,
   0.83{\pm}0.01,
   & \label{wave06} \\
   &
   A_{2}^{{\Upsilon}{\to}B_{c}}(0)\, =\,
   0.73{\pm}0.08,
   & \label{wave07} \\
   &
   V^{{\Upsilon}{\to}B_{c}}(0)\, =\,
   1.98{\pm}0.06,
   & \label{wave08}
   \end{eqnarray}
  where the uncertainties come from variation of
  valence quark mass $m_{b,c}$.
  In addition, according to the NRQCD
  argument, the relativistic corrections and
  higher-twist effects might
  give uncertainties of ${\cal O}(v^{2})$,
  about 10\%${\sim}$30\%.
  Values at $q^{2}$ $\neq$ $0$ could, in principle,
  be extrapolated by assuming the form factors
  dominated by a proper pole which is unknown,
  or calculated with other method, such as the
  approach using the Bethe-Salpeter wave functions
  with the help of the nonrelativistic instantaneous
  approximation and the potential model based
  on the Mandelstam formalism \cite{prd49} and so on.
  Here, we will follow the common practice for
  nonleptonic $B$ decays with the QCDF approach.
  Values of form factors at $q^{2}$ $=$ $0$ are taken
  to offer an order of magnitude estimation,
  because both ${\Upsilon}$ and $B_{c}$ are heavy
  quarkonium and the recoil effects might be not
  so significant.

  \subsection{Decay amplitudes}
  \label{sec0204}
  With the aforementioned definition of hadronic matrix elements,
  the decay amplitudes of ${\Upsilon}(1S)$ ${\to}$ $B_{c}M$
  decays can be written as
    \begin{eqnarray}
  {\cal A}({\Upsilon}{\to}B_{c}^{+}{\pi}^{-})
  &=&
   \sqrt{2}\, G_{F}\, V_{cb}\, V_{ud}^{\ast}\, a_{1}\,
   f_{\pi}\, m_{\Upsilon}\, ({\epsilon}_{\Upsilon}{\cdot}p_{\pi})\,
    A_{0}^{{\Upsilon}{\to}B_{c}}
   \label{amp-bc-pi}, \\
  {\cal A}({\Upsilon}{\to}B_{c}^{+}K^{-})
  &=&
   \sqrt{2}\, G_{F}\, V_{cb}\, V_{us}^{\ast}\, a_{1}\,
   f_{K}\, m_{\Upsilon}\, ({\epsilon}_{\Upsilon}{\cdot}p_{K})\,
   A_{0}^{{\Upsilon}{\to}B_{c}}
   \label{amp-bc-k}, \\
  {\cal A}({\Upsilon}{\to}B_{c}^{+}{\rho}^{-})
  &=&
   -i\,\frac{G_{F}}{\sqrt{2}}\, V_{cb}\, V_{ud}^{\ast}\,
   a_{1}\, f_{\rho}\, m_{\rho}\, \Big\{
   ({\epsilon}_{\Upsilon}{\cdot}{\epsilon}_{\rho}^{\ast})\,
   (m_{\Upsilon}+m_{B_{c}})\, A_{1}^{{\Upsilon}{\to}B_{c}}
   \nonumber \\ & & \!\!\!\!\!\!\!\!\!\!\!\!\!\!\!\!\!\!\!\!
   + ({\epsilon}_{\Upsilon}{\cdot}p_{\rho})\,
     ({\epsilon}_{\rho}^{\ast}{\cdot}p_{\Upsilon})\,
       \frac{ 2\, A_{2}^{{\Upsilon}{\to}B_{c}} }
            { m_{\Upsilon}+m_{B_{c}} }
  -i\,{\epsilon}_{{\mu}{\nu}{\alpha}{\beta}}\,
      {\epsilon}_{\Upsilon}^{\mu}\,
      {\epsilon}_{\rho}^{{\ast}{\nu}}\,
      p_{\Upsilon}^{\alpha}\,p_{\rho}^{\beta}\,
       \frac{2\, V^{{\Upsilon}{\to}B_{c}} }
            { m_{\Upsilon}+m_{B_{c}} } \Big\}
   \label{amp-bc-rho}, \\
  {\cal A}({\Upsilon}{\to}B_{c}^{+}K^{{\ast}-})
  &=&
   -i\,\frac{G_{F}}{\sqrt{2}}\, V_{cb}\, V_{us}^{\ast}\,
   a_{1}\, f_{K^{\ast}}\, m_{K^{\ast}}\, \Big\{
   ({\epsilon}_{\Upsilon}{\cdot}{\epsilon}_{K^{\ast}}^{\ast})\,
   (m_{\Upsilon}+m_{B_{c}})\, A_{1}^{{\Upsilon}{\to}B_{c}}
   \nonumber \\ & & \!\!\!\!\!\!\!\!\!\!\!\!\!\!\!\!\!\!\!\!
   + ({\epsilon}_{\Upsilon}{\cdot}p_{K^{\ast}})\,
     ({\epsilon}_{K^{\ast}}^{\ast}{\cdot}p_{\Upsilon})\,
       \frac{ 2\, A_{2}^{{\Upsilon}{\to}B_{c}} }
            { m_{\Upsilon}+m_{B_{c}} }
  -i\,{\epsilon}_{{\mu}{\nu}{\alpha}{\beta}}\,
      {\epsilon}_{\Upsilon}^{\mu}\,
      {\epsilon}_{K^{\ast}}^{{\ast}{\nu}}\,
      p_{\Upsilon}^{\alpha}\,p_{K^{\ast}}^{\beta}\,
       \frac{2\, V^{{\Upsilon}{\to}B_{c}} }
            { m_{\Upsilon}+m_{B_{c}} } \Big\}
   \label{amp-bc-kv}.
  \end{eqnarray}

  For the ${\Upsilon}(1S)$ ${\to}$ $B_{c}V$ decays,
  the hadronic matrix elements in Eq.(\ref{amp})
  can also be expressed as \cite{vv}
    \begin{eqnarray}
   {\cal H}_{\lambda} & =&
   {\langle}V{\vert}J^{\mu}{\vert}0{\rangle}
   {\langle}B_{c}{\vert}J_{\mu}{\vert}{\Upsilon}(1S){\rangle}
    \nonumber \\ &=&
   {\epsilon}_{V}^{{\ast}{\mu}} {\epsilon}_{{\Upsilon}}^{\nu} \Big\{
    a\,g_{{\mu}{\nu}}
    +\frac{ b  }{m_{{\Upsilon}}\,m_{V}} ( p_{{\Upsilon}} + p_{B_{c}} )^{\mu} p_{V}^{\nu}
    +\frac{i\,c }{m_{{\Upsilon}}\,m_{V}} {\epsilon}_{{\mu}{\nu}{\alpha}{\beta}}
     p_{V}^{\alpha}(p_{{\Upsilon}}+p_{B_{c}})^{\beta} \Big\}
    \label{spd}.
    \end{eqnarray}

  The definition of helicity amplitudes is
    \begin{eqnarray}
   {\cal H}_{0} &=& -a\,y-2b\,(y^{2}-1)
    \label{h0}, \\
   {\cal H}_{\pm} &=& a\, {\pm}\,2c\,\sqrt{y^{2}-1}
    \label{h1},
    \end{eqnarray}
  where invariant amplitudes $a$, $b$, $c$ and variable $y$ are
    \begin{eqnarray}
    a &=& -i\, f_{V}\, m_{V}\, ( m_{{\Upsilon}}+m_{B_{c}} )\,
    A_{1}^{{\Upsilon}{\to}B_{c}}(q^{2})
    \label{sa}, \\
    b &=& -i\, f_{V}\, m_{{\Upsilon}}\, m_{V}^{2}\,
    \frac{A_{2}^{{\Upsilon}{\to}B_{c}}(q^{2})}{m_{{\Upsilon}}+m_{B_{c}}}
    \label{db}, \\
   c &=& +i\, f_{V}\, m_{{\Upsilon}}\, m_{V}^{2}\,
   \frac{V^{{\Upsilon}{\to}B_{c}}(q^{2})}{m_{{\Upsilon}}+m_{B_{c}}}
    \label{pc}, \\
    y &=& \frac{p_{{\Upsilon}}{\cdot}p_{V}}{m_{{\Upsilon}}\,m_{V}}
      \ =\ \frac{m_{{\Upsilon}}^{2}-m_{B_{c}}^{2}+m_{V}^{2}}{2\,m_{{\Upsilon}}\,m_{V}}
    \label{xx}.
    \end{eqnarray}
  The scalar amplitudes $a$, $b$, $c$ describe the $s$, $d$, $p$
  wave contributions, respectively.
  Clearly, compared with the $s$ wave amplitude,
  the $p$ and $d$ wave amplitudes are suppressed
  by a factor $m_{V}/m_{{\Upsilon}}$.

  \section{Numerical results and discussion}
  \label{sec03}

  In the rest frame of ${\Upsilon}(1S)$ particle, branching ratio
  for nonleptonic ${\Upsilon}(1S)$ ${\to}$ $B_{c}M$ weak decays can
  be written as
   \begin{equation}
  {\cal B}r({\Upsilon}(1S){\to}B_{c}M)\ =\ \frac{1}{12{\pi}}\,
   \frac{p_{\rm cm}}{m_{{\Upsilon}}^{2}{\Gamma}_{{\Upsilon}}}\,
  {\vert}{\cal A}({\Upsilon}(1S){\to}B_{c}M){\vert}^{2}
   \label{br},
   \end{equation}
 where the decay width ${\Gamma}_{\Upsilon}$ $=$
 $54.02{\pm}1.25$ keV \cite{pdg};
 the momentum of final states is
   \begin{equation}
   p_{\rm cm}\ =\
   \frac{ \sqrt{ [m_{\Upsilon}^{2}-(m_{B_{c}}+m_{M})^{2}]
                 [m_{\Upsilon}^{2}-(m_{B_{c}}-m_{M})^{2}] }  }
       { 2\,m_{\Upsilon} }
   \label{pcm}.
   \end{equation}

 The input parameters,
 including the CKM Wolfenstein parameters,
 masses of $b$ and $c$ quarks, decay constants,
 and Gegenbauer moments of distribution amplitudes
 in Eq.(\ref{twist}),
 are collected in Table \ref{input}.
 If not specified explicitly, we will take their central
 values as the default inputs.
 Our numerical results on the $CP$-averaged branching ratios
 for the ${\Upsilon}(1S)$ ${\to}$ $B_{c}M$ decays are
 displayed in Table \ref{tabbr}, where theoretical uncertainties
 of the QCDF results come from the CKM parameters,
 the renormalization scale ${\mu}$ $=$ $(1{\pm}0.5)m_{b}$,
 masses of $b$ and $c$ quarks, hadronic parameters
 (decay constants and Gegenbauer moments), respectively.
 For the sake of comparison, previous results of
 Refs. \cite{ijma14,adv2013} are re-evaluated with
 coefficient $a_{1}$ $=$ $1.054$,
 where the scenario of the flavor dependent parameter
 ${\omega}$ in Ref. \cite{adv2013} is taken.
 The following are some comments.

 (1)
 The QCDF's results fall in between those of Ref. \cite{ijma14}
 and Ref. \cite{adv2013}, because the form factors
 $A_{0,1}^{{\Upsilon}{\to}B_{c}}$ in our calculation fall in
 between those of Ref. \cite{ijma14} and Ref. \cite{adv2013}.

 (2) There is a clear hierarchical relationship,
 ${\cal B}r({\Upsilon}(1S){\to}B_{c}{\rho})$ $>$
 ${\cal B}r({\Upsilon}(1S){\to}B_{c}{\pi})$ $>$
 ${\cal B}r({\Upsilon}(1S){\to}B_{c}K^{\ast})$ $>$
 ${\cal B}r({\Upsilon}(1S){\to}B_{c}K)$.
 These are two dynamical reasons.
 One is that the CKM factor $V_{cb}V_{us}^{\ast}$
 responsible for the
 ${\Upsilon}(1S)$ ${\to}$ $B_{c}K^{(\ast)}$ decay
 is suppressed by a factor of ${\lambda}$ relative to
 the CKM factor $V_{cb}V_{ud}^{\ast}$ responsible for the
 ${\Upsilon}(1S)$ ${\to}$ $B_{c}{\pi}$, $B_{c}{\rho}$ decays.
 The other is that the orbital angular momentum $L_{B_{c}P}$
 $>$ $L_{B_{c}V}$.

 (3)
 The CKM-favored $a_{1}$ dominated ${\Upsilon}(1S)$
 ${\to}$ $B_{c}{\rho}$ decay has the largest branching
 ratio, ${\sim}$ $10^{-10}$,
 which should be sought for with high priority
 and firstly observed at the running LHC and
 forthcoming SuperKEKB.
 For example, the ${\Upsilon}(1S)$ production
 cross section in p-Pb collision can reach up
 to a few ${\mu}b$ with the ALICE detector at
 LHC \cite{plb740}.
 Therefore, per 100 $fb^{-1}$ data collected at
 ALICE, over $10^{11}$ ${\Upsilon}(1S)$ particles
 are in principle available, corresponding to
 tens of ${\Upsilon}(1S)$ ${\to}$ $B_{c}{\rho}$
 events if with about 10\% reconstruction efficiency.

 (4)
 There are many uncertainties on the QCDF's results.
 The first uncertainty, about 7${\sim}$8\%, from
 the CKM factors could be lessened with the
 improvement on the precision of the Wolfenstein
 parameter $A$.
 The second uncertainty from the renormalization scale
 should, in principle, be reduced by inclusion of higher
 order ${\alpha}_{s}$ corrections to hadronic matrix
 elements.
 The third uncertainty is due to the fact that
 masses of $b$ and $c$ quark affect the shape lines
 of wave functions, and hence the magnitude of form
 factors and branching ratios.
 The fourth uncertainty from hadronic parameters is expected
 to be reduced with the relative ratio of branching ratios.

 (5)
 Other factors, such as the contributions
 of higher order corrections to hadronic matrix
 elements, relativistic effects, $q^{2}$ dependence of
 form factors and so on,
 which are not considered in this paper,
 deserve the dedicated study.
 Our results just provide an order of magnitude estimation.

   \begin{table}[ht]
   \caption{The values of input parameters.}
   \label{input}
   \begin{ruledtabular}
  \begin{tabular}{ll}
  \multicolumn{2}{c}{Wolfenstein parameters} \\ \hline
    ${\lambda}$  $=$ $0.22537{\pm}0.00061$     \cite{pdg}
  & $A$          $=$ $0.814^{+0.023}_{-0.024}$ \cite{pdg} \\ \hline
    \multicolumn{2}{c}{masses of quarks} \\ \hline
    $m_{c}$ $=$ $1.67{\pm}0.07$ GeV  \cite{pdg}
  & $m_{b}$ $=$ $4.78{\pm}0.06$ GeV  \cite{pdg} \\ \hline
  \multicolumn{2}{c}{decay constants} \\ \hline
    $f_{\pi}$ $=$ $130.41{\pm}0.20$ MeV \cite{pdg}
  & $f_{K}  $ $=$ $156.2{\pm}0.7$ MeV \cite{pdg} \\
    $f_{\rho}$   $=$ $216{\pm}3$ MeV \cite{ballv}
  & $f_{K^{\ast}}$ $=$ $220{\pm}5$ MeV \cite{ballv} \\ \hline
  \multicolumn{2}{c}{Gegenbauer moments at the scale ${\mu}$ $=$ 1 GeV} \\ \hline
    $a_{1}^{\rho}$ $=$ $0$ \cite{ballv}
  & $a_{2}^{\rho}$ $=$ $0.15{\pm}0.07$ \cite{ballv} \\
    $a_{1}^{K^{\ast}}$ $=$ $-0.03{\pm}0.02$ \cite{ballv}
  & $a_{2}^{K^{\ast}}$ $=$ $0.11{\pm}0.09$ \cite{ballv} \\
    $a_{1}^{\pi}$ $=$ $0$ \cite{ballp}
  & $a_{2}^{\pi}$ $=$ $0.25{\pm}0.15$ \cite{ballp} \\
    $a_{1}^{K}$ $=$ $-0.06{\pm}0.03$ \cite{ballp}
  & $a_{2}^{K}$ $=$ $0.25{\pm}0.15$ \cite{ballp}
  \end{tabular}
  \end{ruledtabular}
  \end{table}
   \begin{table}[h]
   \caption{The $CP$-averaged branching ratios for the
   ${\Upsilon}(1S)$ ${\to}$ $B_{c}M$ decays.}
   \label{tabbr}
   \begin{ruledtabular}
  \begin{tabular}{l|c|c|c}
    & Ref. \cite{ijma14} & Ref. \cite{adv2013} & this work \\ \hline
    $10^{10}{\times}{\cal B}r({\Upsilon}(1S){\to}B_{c}{\rho})$
  & $1.84$
  & $1.2$
  & $1.53^{+0.11+0.10+0.03+0.04}_{-0.10-0.04-0.04-0.04}$ \\ \hline
    $10^{11}{\times}{\cal B}r({\Upsilon}(1S){\to}B_{c}{\pi})$
  & $6.91$
  & $2.8$
  & $5.03^{+0.36+0.34+0.09+0.02}_{-0.34-0.14-0.11-0.02}$ \\ \hline
    $10^{12}{\times}{\cal B}r({\Upsilon}(1S){\to}B_{c}K^{\ast})$
  & $10.47$
  & $6.2$
  & $8.75^{+0.68+0.55+0.16+0.40}_{-0.64-0.20-0.22-0.39}$ \\ \hline
    $10^{12}{\times}{\cal B}r({\Upsilon}(1S){\to}B_{c}K)$
  & $5.03$
  & $2.3$
  & $3.73^{+0.29+0.25+0.07+0.03}_{-0.27-0.10-0.08-0.03}$
  \end{tabular}
  \end{ruledtabular}
  \end{table}

  \section{Summary}
  \label{sec04}
  With the sharp increase of the ${\Upsilon}(1S)$ data sample
  at high-luminosity dedicated heavy-flavor factories,
  the bottom-changing ${\Upsilon}(1S)$ ${\to}$ $B_{c}M$ weak decays
  are interesting in exploring the underlying mechanism responsible
  for transition between heavy quarknoia, investigating perturbative
  and nonperturbative effects and overconstraining parameters from
  $B$ decays.
  The ${\Upsilon}(1S)$ weak decays are allowable within
  the standard model, though their branching ratios are expected
  to be tiny in comparison to the conventional strong and
  electromagnetic decays.
  In this paper, we studied the nonleptonic ${\Upsilon}(1S)$
  ${\to}$ $B_{c}M$ weak decays, which are $a_{1}$-dominated
  based on the low energy effective theory, and hence should
  have large branching ratios among weak decay modes.
  Considering the nonfactorizable contributions to hadronic
  matrix elements with the QCDF approach, we estimated the
  branching ratios of the ${\Upsilon}(1S)$ ${\to}$ $B_{c}M$
  weak decays, where transition form factors are obtained
  by the integrals of wave functions with the nonrelativistic
  isotropic harmonic oscillator potential.
  The prediction on branching ratios for the ${\Upsilon}(1S)$
  ${\to}$ $B_{c}M$ decays is the same order as previous
  works \cite{ijma14,adv2013}.
  The CKM favored ${\Upsilon}(1S)$ ${\to}$ $B_{c}{\rho}$ decay
  has relatively large branching ratio, ${\sim}10^{-10}$,
  and might be detectable in future experiments.

  \section*{Acknowledgments}
  We thank the referees for their helpful comments.
  The work is supported by the National Natural Science Foundation
  of China (Grant Nos. 11475055, 11275057, U1232101 and U1332103).
  Dr. Wang thanks for the support from CCNU-QLPL Innovation Fund (QLPL201411).

  

\begin{thebibliography}{99}
  \bibitem{herb}
          S. Herb {\em et al.}, Phys. Rev. Lett. 39, 252 (1977).
  \bibitem{innes}
          W. Innes {\em et al.}, Phys. Rev. Lett. 39, 1240 (1977).
  \bibitem{ann1983}
          P. Franzini, J. Lee-Franzini, Ann. Rev. Nucl. Part. Sci. 33, 1 (1983).
  \bibitem{pdg}
          K. Olive {\em et al.} (Particle Data Group), Chin. Phys. C 38, 090001 (2014).
  \bibitem{ozi-o}
          S. Okubo, Phys. Lett. 5, 165 (1963).
  \bibitem{ozi-z}
          G. Zweig, CERN-TH-401, 402, 412 (1964).
  \bibitem{ozi-i}
          J. Iizuka, Prog. Theor. Phys. Suppl. 37-38, 21 (1966).
  \bibitem{0807.1427}
          W. Love {\em et al.} (CLEO Collaboration), Phys. Rev. Lett. 101, 151802 (2008).
  \bibitem{pos2013.higgs}
          M. Chang, {\em et al.} (Belle Collaboration), PoS EPS-HEP2013, 270 (2013).
  \bibitem{1502.06019}
          J. Lees {\em et al.} (BaBar Collaboration), Phys. Rev. D. 91, 071102 (2015).
  \bibitem{1212.6552}
          C. Patrignani, T. Pedlar, J. Rosner, Annu. Rev. Nucl. Part. Sci. 63, 21 (2013).
  \bibitem{qcdf1}
          M. Beneke {\em et al.}, Phys. Rev. Lett. 83, 1914 (1999).
  \bibitem{qcdf2}
          M. Beneke {\em et al.}, Nucl. Phys. B 591, 313 (2000).
  \bibitem{1212.5342}
          J. Brodzicka {\em et al.} (Belle Collaboration), Prog. Theor. Exp. Phys. 2012, 04D001.
  \bibitem{1108.5874}
          J. Lees {\em et al.} (BaBar Collaboration), Phys. Rev. D 84, 092003 (2011).
  \bibitem{0704.2766}
          R. Briere {\em et al.} (CLEO Collaboration), Phys. Rev. D 76, 012005 (2007).
  \bibitem{1403.3648}
          B. Abelev {\em et al.} (ALICE Collaboration), Eur. Phys. J. C 74, 2974 (2014).
  \bibitem{1212.7255}
          G. Ada {\em et al.} (ATLAS Collaboration), Phys. Rev. D 87, 052004 (2013).
  \bibitem{1501.07750}
          V. Khachatryan {\em et al.} (CMS Collaboration), arXiv:1501.07750.
  \bibitem{1304.6977}
          R. Aaij {\em et al.} (LHCb Collaboration), JHEP 1306, 064, (2013).
  \bibitem{zpc62.271}
          M. Sanchis-Lozano, Z. Phys. C 62, 271 (1994).
  \bibitem{ijma14}
          K. Sharma, R. Verma, Int. J. Mod. Phys. A 14,  937 (1999).
  \bibitem{adv2013}
          R. Dhir, R. Verma, A. Sharma, Advances in High Energy Physics, 2013, 706543 (2013).
  \bibitem{bsw1}
          M. Wirbel, B. Stech, M. Bauer, Z. Phys. C 29, 637 (1985).
  \bibitem{prd46}
          G. Legage {\em et al.}, Phys. Rev. D 46, 4052 (1992).
  \bibitem{prd51}
          G. Bodwin, E. Braaten, G. Legage, Phys. Rev. D 51, 1125 (1995).
  \bibitem{rmp77}
          N. Brambilla {\em et al.}, Rev. Mod. Phys. 77, 1423 (2005).
  \bibitem{9512380}
          G. Buchalla, A. Buras, M. Lautenbacher, Rev. Mod. Phys. 68, 1125, (1996).
  \bibitem{bjorken}
          J. Bjorken, Nucl. Phys. B (Proc. Suppl.) 11, 325 (1989).
  \bibitem{nf1}
          N. Cabibbo, L. Maiani, Phys. Lett. B 73, 418 (1978).
  \bibitem{nf2}
          D. Fakirov, B. Stech, Nucl. Phys. B 133, 315 (1978).
  \bibitem{ballv}
          P. Ball and G. Jones, JHEP 03, 069 (2007).
  \bibitem{ballp}
          P. Ball, V. Braun, A. Lenz, JHEP 05, 004 (2006).
  \bibitem{prd77}
          J. Sun {\em et al.}, Phys. Rev. D 77, 074013 (2008).
  \bibitem{xiao}
          B. Xiao, X. Qin, B. Ma, Eur. Phys. J. A 15, 523 (2002).
  \bibitem{prd49}
          C. Chang and Y. Chen, Phys. Rev. D 49, 3399 (1994).
  \bibitem{vv}
          G. Valencia, Phys. Rev. D 39, 3339 (1989).
  \bibitem{plb740}
          B. Abelev, Phys. Lett. B 740, 105 (2015).
  \end{thebibliography}
  \end{document}